\documentclass[aps,prl,reprint]{revtex4-1}

\usepackage{natbib}
\usepackage{graphicx}

\usepackage{graphics}
\usepackage{amssymb}
\usepackage{amsbsy}
\usepackage{amsfonts}
\usepackage{bm}
\usepackage{mathtools}
\usepackage{empheq}

\usepackage{amsmath}

\newcommand{\kpar}{k_\parallel}
\newcommand{\sign}{\textrm{sign}}
\newcommand{\pr}{\partial}
\newcommand{\nn}{\nonumber}

\begin{document}

\title{New closures for more precise modeling \\
  of Landau damping in the fluid framework} 


\author{P. Hunana}
\altaffiliation{Center for Space Plasma and Aeronomic Research (CSPAR),
University of Alabama, Huntsville, AL 35805, USA}

\author{G. P. Zank}
\altaffiliation{Center for Space Plasma and Aeronomic Research (CSPAR),
  University of Alabama, Huntsville, AL 35805, USA}

\author{M. Laurenza}
\altaffiliation{National Institute for Astrophysics, Institute for Space Astrophysics and Planetology
  (INAF-IAPS), Rome, 00133, Italy}

\author{A. Tenerani}
\altaffiliation{Department of Earth, Planetary, and Space Sciences, University of California, Los Angeles,
  CA 90095, USA}

\author{\\G. M. Webb}
\altaffiliation{Center for Space Plasma and Aeronomic Research (CSPAR),
  University of Alabama, Huntsville, AL 35805, USA}

\author{M. L. Goldstein}
\altaffiliation{Space Science Institute, Boulder, CO 80301, USA}

\author{M. Velli}
\altaffiliation{Department of Earth, Planetary, and Space Sciences, University of California, Los Angeles,
  CA 90095, USA}

\author{L. Adhikari}
\altaffiliation{Center for Space Plasma and Aeronomic Research (CSPAR),
  University of Alabama, Huntsville, AL 35805, USA}



\begin{abstract}
  Incorporation of kinetic effects such as Landau damping into a fluid framework was pioneered by Hammett and Perkins PRL 1990, by obtaining closures
  of the fluid hierarchy, where the gyrotropic heat flux fluctuations or the deviation of the 4th-order gyrotropic fluid moment, are expressed through lower-order fluid moments.
  To obtain a closure of a fluid model expanded around a bi-Maxwellian distribution function, the usual plasma dispersion function $Z(\zeta)$ that appears in kinetic theory
  or the associated plasma response function $R(\zeta)=1 + \zeta Z(\zeta)$, have to be approximated with a suitable Pad\'e approximant in such a way,
  that the closure is valid for all $\zeta$ values. Such closures are rare, and the original closures of Hammett and Perkins are often employed.
  Here we present a complete mapping of all plausible Landau fluid closures that can be constructed at the level of 4th-order moments in the gyrotropic limit and
  we identify the most precise closures. Furthermore, by considering 1D closures at higher-order moments,
    we show that it is possible to reproduce linear Landau damping in the fluid framework to any desired precision, thus showing convergence of the fluid and collisionless
  kinetic descriptions.
\end{abstract}

\maketitle
Fluid models are an extremely important tool in many areas of space physics and astrophysics.
Despite the underlying dynamics of these systems being often almost completely collisionless,
theoretical models and numerical simulations with
simplified fluid models that \emph{implicitly} assume a high-collisionality regime, such as magnetohydrodynamics (MHD) \cite{Goldstein1995,TuMarsch1995,Zank1999,Zhou2004,BrunoCarbone2013},
provided deep insight into many phenomena, such as the solar wind, the global structure of the heliosphere, turbulence theories, magnetic reconnection
and many others. The implicit assumption of high collisionality in MHD comes from prescribing the pressure to be a scalar
quantity, i.e., by prescribing that the underlying distribution function is strictly isotropic and that it \emph{remains} strictly isotropic during its time evolution.
In collisionless systems, the distribution function is free to evolve from its initial state and become anisotropic, before micro-instabilities start to regulate/restrict
its further anisotropic evolution. In another words, the implicit assumption of high-collisionality in MHD comes from prescribing the pressure
\emph{fluctuations} to be isotropic.
The absence of anisotropic pressure fluctuations in compressible
MHD is the main reason why MHD deviates (even at the linear level for an isotropic Maxwellian), from the simplest collisionless fluid description, known as 
CGL (after Chew, Goldberger and Low \cite{Chew1956,Abraham-Shrauner1967,FerriereAndre2002,HunanaZank2017,Tenerani2017}) and also sometimes referred to as collisionless MHD.  
Nevertheless, even in the low-frequency long-wavelength limit, the CGL fluid model still deviates from a collisionless kinetic description, primarily
because of the absence of the kinetic effect of Landau damping \cite{Landau1946}.
For example, consider a proton-electron plasma with external magnetic field $\boldsymbol{B}_0$, where both species are
described by an equilibrium bi-Maxwellian distribution function, and consider the usual ion-acoustic (sound) mode that propagates in the direction parallel to $\boldsymbol{B}_0$. 
At wavelengths that are much longer than the Debye length, the exact kinetic dispersion relation reads
\begin{equation} \label{eq:soundKin}
\frac{T_{\parallel e}^{(0)}}{T_{\parallel p}^{(0)}} R(\zeta_p) +R(\zeta_e) =0,
\end{equation}
where the plasma response function $R(\zeta)=1 + \zeta Z(\zeta)$ and the plasma dispersion function
$Z(\zeta)=\frac{1}{\sqrt{\pi}} V.P. \int_{-\infty}^\infty\frac{e^{-x^2}}{x-\zeta}dx+i\sqrt{\pi}e^{-\zeta^2}$ $\forall$ $\textrm{Im}(\zeta)$, and the integration passes ``through'' the pole.
With species index $r$, the variable $\zeta_r$ is here defined as $\zeta_r=\omega/(|\kpar|v_{\textrm{th}\parallel r})$, $\omega$ being frequency and $\kpar$ the parallel wavenumber,
the parallel thermal speed $v_{\textrm{th}\parallel r}=\sqrt{2T_{\parallel r}^{(0)}/m_r}$,
and $T_{\parallel r}^{(0)}=p_{\parallel r}^{(0)}/n_r^{(0)}$ is the parallel equilibrium temperature. The dispersion relation (\ref{eq:soundKin}) can in general be solved
only numerically, and for example for $\tau\equiv T_{\parallel e}^{(0)}/T_{\parallel p}^{(0)}=1$, the solution is $\zeta_p=\pm 1.457-0.627i$. The negative imaginary part represents
strong Landau damping, and since no dispersive effects are present, the Landau damping of the parallel ion-acoustic mode does not disappear
even on large astrophysical scales, i.e. in the low-frequency long-wavelength limit where the phase speed $\omega/\kpar$ is constant. In contrast, the
solution for an ion-acoustic mode with both species described by the CGL pressure equations reads $\zeta_p = \pm \sqrt{\frac{3}{2}\frac{(1+\tau)}{(1+\mu)}}$, where
$\mu\equiv m_e/m_p = 1/1836$, so for $\tau=1$ the solution is $\zeta_p=\pm 1.732$. Alternatively, if the electrons are prescribed to be isothermal, the
dispersion relation reads $\zeta_p = \pm \sqrt{\frac{1}{2}\frac{(3+\tau)}{(1+\mu)}}$, which for $\tau=1$ yields $\zeta_p=\pm 1.414$. Therefore,
without Landau damping the usual fluid models do not represent the correct long-wavelength limit of collisionless kinetic theory.

The incorporation of Landau damping into the CGL fluid model was pioneered by Hammett and Perkins \cite{HammettPerkins1990} and was further refined (for example
\cite{Hammett1992,Snyder1997,PassotSulem2007,PSH2012,SulemPassot2015} and references therein). These fluid models that describe Landau damping in the fluid framework
are usually referred to as gyrofluids (formulated in the guiding-center reference frame) or Landau fluids (formulated in the usual laboratory reference frame),
even though there are other subtle differences and the vocabulary is not strictly enforced.
These fluid models are constructed by calculating the hierarchy of fluid moments of the Vlasov equation to
higher-orders than the usual pressure tensor, and by finding a closure, where the last retained fluid moment is expressed through lower-order moments. 
To find a closure, the exact kinetic $R(\zeta)$ function is replaced by a suitable Pad\'e approximant (as a ratio of two polynomials) in such a way, that the closure
is valid for all $\zeta$ values. A (generalized) n-pole Pad\'e approximant $R_n(\zeta)$ to a function $R(\zeta)$ is found by matching the power series
expansion $|\zeta|\ll 1$ and the asymptotic series expansion $|\zeta|\gg 1$ of both functions.
There are of course many possible choices, and here we are interested
only in approximants that at least reproduce the first term of the asymptotic expansion $R(\zeta)=-1/(2\zeta^2)+\cdots$, i.e. as having a precision $o(\zeta^{-2})$.
Here we define ``the basic'' n-pole Pad\'e approximant of $R(\zeta)$ as 
\begin{equation}
 R_{n,0}(\zeta) = \frac{1+a_1 \zeta + a_2\zeta^2+\cdots +a_{n-2}\zeta^{n-2} }{ 1+b_1\zeta+b_2\zeta^2 + \cdots+b_{n-1}\zeta^{n-1} -2a_{n-2} \zeta^n }, \label{eq:Rn0}\nn
\end{equation}
where the second index in $R_{n,n'}(\zeta)$ signifies, that $n'$ \emph{additional} asymptotic points were
used in comparison with the basic $R_{n,0}(\zeta)$ definition. 
The $n'=0$ index helps to quickly orient a large hierarchy of many possible $R(\zeta)$ approximants.
This asymptotic profile correctly captures the asymptotic decay of the density moment,
and any profile with fewer asymptotic points should be avoided if possible. The 1-pole approximant is $R_1(\zeta)=1/(1-i\sqrt{\pi}\zeta)$.
$R_{n,0}(\zeta)$ has power series precision $o(\zeta^{2n-3})$ and asymptotic series precision $o(\zeta^{-2})$, so $R_{n,n'}(\zeta)$ has precision
$o(\zeta^{2n-3-n'})$ and $o(\zeta^{-2-n'})$. The Pad\'e approximant to $Z(\zeta)$ is
defined as $R_{n,n'}(\zeta)=1+\zeta Z_{n,n'}(\zeta)$. Comparison with the 2-index notation of Mart\'in et al. \cite{Martin1980} (introducing superscript M) and of
Hedrick and Leboeuf \cite{HedrickLeboeuf1992} (superscript HL) can be done easily according to $Z_{n,n'}^M = Z_{\frac{n+n'}{2}, n'-3}$ and $Z_{n,n'}^{HL} = Z_{n,n'+n-3}$.
Pad\'e approximants were also used in developing analytic models for the Rayleigh-Taylor and Richtmyer-Meshkov instability \cite{Zhou2017-1,Zhou2017-2}.

Similarly to \cite{HammettPerkins1990}, we concentrate here on a 1D geometry that can be viewed as an electrostatic case, or from our view preferably
as propagation along $B_0$, which naturally picks up the ion-acoustic mode (since the 1D velocity fluctuations are along $B_0$). For brevity we drop writing the
parallel subscripts (except on $\kpar$) and species index $r$,  since closures are constructed independently for each species.
Examples of $R(\zeta)$ Pad\'e approximants are $R_{2,0}(\zeta)=1/(1-i\sqrt{\pi}\zeta-2\zeta^2)$,
\begin{equation}
  R_{3,0}(\zeta) = \frac{1-i\sqrt{\pi}\frac{\pi-3}{4-\pi}\zeta}{1-i\frac{\sqrt{\pi}}{4-\pi}\zeta -\frac{3\pi-8}{4-\pi}\zeta^2 +2i\sqrt{\pi}\frac{\pi-3}{4-\pi}\zeta^3 };\nn
\end{equation}  
$R_{3,1}(\zeta) = \frac{1-i\frac{4-\pi}{\sqrt{\pi}}\zeta}{1-\frac{4i}{\sqrt{\pi}}\zeta-2\zeta^2+2i\frac{4-\pi}{\sqrt{\pi}}\zeta^3}$.
We note that Table 1 of \cite{HedrickLeboeuf1992} can be recovered analytically, and we report a typo
in their $a_1$ coefficient for $Z_{3,1}(\zeta)$ that should be $a_1=\frac{2}{4-\pi}=2.32990$ instead of $2.23990$, used for example in \cite{PassotSulem2007}. 
The two Pad\'e approximants used by \cite{HammettPerkins1990} read
$R_{3,2}(\zeta) = \frac{1-\frac{i\sqrt{\pi}}{2}\zeta}{1-\frac{3i\sqrt{\pi}}{2}\zeta-2\zeta^2+i\sqrt{\pi} \zeta^3}$;
\begin{equation}
 R_{4,3}(\zeta) = \frac{1-i\frac{\sqrt{\pi}}{2}\zeta-\frac{(3\pi-8)}{4}\zeta^2}{1-i\frac{3\sqrt{\pi}}{2}\zeta-\frac{(9\pi-16)}{4}\zeta^2
    +i\sqrt{\pi}\zeta^3+\frac{(3\pi-8)}{2}\zeta^4}\nn,
\end{equation}  
where the first choice yields a closure for the heat flux $q^{(1)} = -i \frac{2}{\sqrt{\pi}} n_0 v_{\textrm{th}} \sign(\kpar)T^{(1)}$. Note that our definition of the thermal
speed contains a factor of 2. The second choice yields a closure for $\widetilde{r}$ defined as $r=3p^2/\rho+\widetilde{r}$ where the 4-th order moment $r=m\int (v-u)^4 f d^3v$
(we follow the notation of \cite{PassotSulem2007}; $\widetilde{r}$ can be also denoted as $\delta r$) and the  
$R_{4,3}(\zeta)$ closure obtained by \cite{HammettPerkins1990} reads
\begin{eqnarray} \label{eq:Static_R43}
  \widetilde{r}^{(1)} = -\frac{i2\sqrt{\pi}}{(3\pi-8)}v_{\textrm{th}} \sign(k_\parallel) q^{(1)}
  + \frac{(32-9\pi)}{2(3\pi-8)}v_{\textrm{th}}^2 n_{0} T^{(1)}. \nn
\end{eqnarray}
Curiously, it can be shown that the fluid dispersion relation that uses the above closure, is 
equivalent to the kinetic dispersion relation (1) once the exact $R(\zeta)$ is replaced by the approximant $R_{4,3}(\zeta)$ (strictly speaking it is equivalent to the
numerator of (1) once both terms in (1) are written with common denominator). Electron inertia must be considered and the displacement current must of course be neglected
in the fluid model to yield (1). This observation is also true for all other $R_{n,n'}(\zeta)$ closures presented here 
and closures that satisfy (1) can be viewed as ``reliable'' or physically-meaningful.
\begin{figure*}[!htpb]
\centering \includegraphics[width=0.7\textwidth]{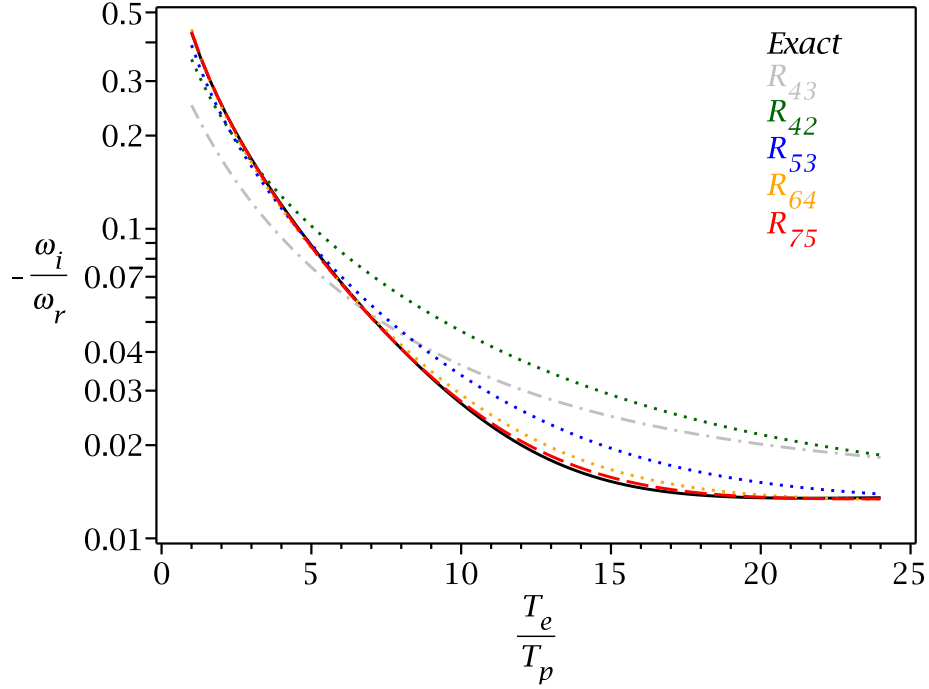}
\caption{ Landau damping of the ion-acoustic mode, calculated with exact $R(\zeta)$ - black solid line; $R_{4,2}(\zeta)$ - green dotted line;
    $R_{5,3}(\zeta$) - blue dotted line;
    $R_{6,4}(\zeta)$ - orange dotted line; and $R_{7,5}(\zeta)$ - red dashed line. The x-axis is the ratio of electron and proton temperature and the y-axis the ratio of
    the damping and real frequency. The solutions represent the most precise dynamic closures that can be
    constructed for the 3rd, 4th, 5th and 6th-order fluid moments.  The $R_{4,3}(\zeta)$ closure
    of \cite{HammettPerkins1990} is plotted as a gray dot-dashed line.
    The figure shows that it is possible to reproduce Landau damping in the fluid framework to any desired precision.} 
\end{figure*}

In Figure 1, the dispersion relation of the fluid model that uses the above $R_{4,3}(\zeta)$ closure (gray dot-dashed line) is compared to the
exact kinetic solution (1) (black solid line).
The figure is motivated by Figure 9.18 in \cite{GurnettBhattacharjee2005} (page 355).
A closure is called ``static'' when the last retained moment (i.e. $\widetilde{r}$) is directly expressed through lower
order moments. A closure is called ``time-dependent'' or ``dynamic'', when the closure involves $\pr/\pr t$ of the last retained moment (i.e. $\zeta\widetilde{r}$), and
the $\pr/\pr t$ is then replaced by a $d/dt$ to recover the Galilean invariance.
Time-dependent closures can be constructed usually with a higher-order Pad\'e approximant than static closures, however,
the replacement of $\pr/\pr t$ with $d/dt$ introduces only one nonlinearity among other neglected nonlinearities.

Here we report on the most precise Landau fluid closures that can be constructed at a given level.
For example, by using $R_{3,1}(\zeta)$, the following static closure can be constructed for the heat flux
\begin{equation}
q^{(1)} = \frac{3\pi-8}{4-\pi}n_0 T^{(0)} u^{(1)}-i\frac{\sqrt{\pi}}{4-\pi}n_0 v_{\textrm{th}}\sign(\kpar) T^{(1)}.
\end{equation}
Considering \emph{power series precision (henceforth abbreviated as p.s.p.)}, this is the most precise static closure that can be constructed for the heat flux,
  and the precision is $o(\zeta^2)$.
The coefficients of the $R_{4,2}(\zeta)$ approximant are $b_3=-2a_1$; $b_2=3a_2-2$;
$a_1 = -i\frac{\sqrt{\pi}(10-3\pi)}{(3\pi-8)}$; $a_2 = -\frac{(16-5\pi)}{(3\pi-8)}$;
$b_1 = -i \frac{2 \sqrt{\pi}}{(3\pi-8)}$,
and the static closure with the highest p.s.p., $o(\zeta^3)$, that can be constructed at the 4th-moment level reads
\begin{eqnarray} \label{eq:Static_R42}
  \widetilde{r}^{(1)} &=& -i\sqrt{\pi} \frac{(10-3\pi)}{(16-5\pi)} v_{\textrm{th}} \sign(k_\parallel) q^{(1)} \nn\\
 && + \frac{(21\pi-64)}{2(16-5\pi)}v_{\textrm{th}}^2 n_{0} T^{(1)} \nn\\
  && + i\sqrt{\pi} \frac{(9\pi-28)}{(16-5\pi)}v_{\textrm{th}} T^{(0)} n_{0}\sign(k_\parallel) u^{(1)}.
\end{eqnarray}
The $R_{4,2}(\zeta)$ is also used to obtain the dynamic closure for the heat flux with the highest p.s.p., and written for a change in real space,
the closure reads
\begin{eqnarray}
&&     \Big[ \frac{d}{d t} -\sqrt{\pi}\frac{10-3\pi}{16-5\pi} v_{\textrm{th}} \pr_z\mathcal{H} \Big] q^{(1)}
  = -n_0 v_{\textrm{th}}^2 \frac{3\pi-8}{16-5\pi} \pr_z T^{(1)} \nn\\
&&\quad  -n_0 T^{(0)}v_{\textrm{th}} \sqrt{\pi}\frac{9\pi-28}{16-5\pi} \pr_z\mathcal{H} u^{(1)}. \label{eq:Dynamic_R42}
\end{eqnarray}
The $\mathcal{H}$ operator is the negative Hilbert transform operator that acts on a function $f(z)$ according to
$\mathcal{H} f(z) \equiv -\frac{1}{\pi z} * f(z) \equiv - \frac{1}{\pi} V.P. \int_{-\infty}^\infty \frac{f(z')}{z-z'} dz'$,
the $*$ operator being the convolution. We use the Fourier decomposition $e^{-i\omega t +i\kpar z}$, and the transformation of a closure between
Fourier and real space can be done simply according to
$-i\omega\leftrightarrow \pr/\pr t$; $i\kpar\leftrightarrow \pr_z$; $i\sign(\kpar)\leftrightarrow \mathcal{H}$, and $|\kpar|\leftrightarrow -\pr_z\mathcal{H}$.
The closure is plotted in Figure 1 as a dark green dotted line and the closure is very accurate in the region $\tau=[1,5]$.

A closure that has the highest p.s.p. at the 4th-moment level, $o(\zeta^4)$,
is a dynamic closure constructed with approximant $R_{5,3}(\zeta)$, that has coefficients $b_5=-2a_3$; $b_4=-2a_2$; $b_3=3a_3-2a_1$; $b_2=3a_2-2$, 
  $a_1 = \frac{i}{\sqrt{\pi}}\frac{(27\pi^2-126\pi+128)}{3(9\pi-28)}$; $a_2 = \frac{(33\pi-104)}{3(9\pi-28)}$;
  $a_3 = \frac{i}{\sqrt{\pi}} \frac{2(9\pi^2-69\pi+128)}{3(9\pi-28)}$; $b_1 = -\frac{i}{\sqrt{\pi}}\frac{2(21\pi-64)}{3(9\pi-28)}$,
and the closure reads
\begin{eqnarray}
&& \Big[ \frac{d}{dt} -\frac{(104-33\pi)\sqrt{\pi}}{2(9\pi^2-69\pi+128)} v_{\textrm{th}}\partial_z\mathcal{H} \Big] \widetilde{r}^{(1)} \nn\\
&&  = v_{\textrm{th}}^2 n_0 T^{(0)}\frac{(135\pi^2-750\pi+1024)}{2(9\pi^2-69\pi+128)} \partial_z u^{(1)} \nn\\
&& \quad + n_0 v_{\textrm{th}}^3 \frac{3(160-51\pi)\sqrt{\pi}}{4(9\pi^2-69\pi+128)} \partial_z\mathcal{H} T^{(1)} \nn\\ 
&& \quad + v_{\textrm{th}}^2 \frac{(54\pi^2-333\pi+512)}{2(9\pi^2-69\pi+128)} \partial_z q^{(1)}. \label{eq:R53_closure}
\end{eqnarray}
The dispersion relation of a fluid model that uses the $R_{5,3}(\zeta)$ closure is plotted in Figure 1 as a blue dotted line.
In the region $\tau=[1,5]$, this is the most precise closure that can be constructed at the 4th-moment level. 

In contrast, a static closure that uses the most asymptotic series $|\zeta|\gg 1$ points at the 4th-moment level,
with precision $o(\zeta^{-6})$, is constructed with $R_{4,4}(\zeta)$,
and the closure reads $\widetilde{r}^{(1)} = -\frac{3}{4}\sqrt{\pi} v_{\textrm{th}} \mathcal{H} q^{(1)}$.
The most asymptotically precise closure is a dynamic closure constructed with $R_{5,6}(\zeta)$,
that has a precision $o(\zeta^{-8})$ and the closure reads 
$\big[ \frac{d}{dt} - \frac{8}{3\sqrt{\pi}} v_{\textrm{th}} \pr_z\mathcal{H} \big]\widetilde{r}^{(1)} = -2v_{\textrm{th}}^2 \pr_z q^{(1)}$.
For temperatures $\tau=[15,100]$, this is the most precise closure that can be constructed at the 4th-moment level.

We mapped all the possible Landau fluid closures that can be constructed (at the level of heat flux or the moment $\widetilde{r}$)  
and there are 7 possible static closures (5 reliable), and 13 dynamic closures (9 reliable), some of them related. We do not provide analytic
solutions for all of these closures.
Nevertheless, other notable closures are for $R_{5,4}(\zeta)$:
$\big[ \frac{d}{dt} - \frac{21\pi-64}{\sqrt{\pi}(9\pi-28)} v_{\textrm{th}} \pr_z\mathcal{H} \big] \widetilde{r}^{(1)} 
  = -n_0v_{\textrm{th}}^3 \frac{256-81\pi}{2(9\pi-28)\sqrt{\pi}} \pr_z\mathcal{H} T^{(1)}
  -v_{\textrm{th}}^2 \frac{32-9\pi}{2(9\pi-28)} \pr_z q^{(1)}$,
and for $R_{5,5}(\zeta)$:
$ \big[ \frac{d}{dt} - \frac{6\sqrt{\pi}}{(32-9\pi)}v_{\textrm{th}} \pr_z\mathcal{H} \big]\widetilde{r}^{(1)}
  = -v_{\textrm{th}}^2 \frac{9\pi}{2(32-9\pi)} \pr_z q^{(1)}.$

All the above closures are also applicable to a 3D geometry when written for $\widetilde{r}_{\parallel\parallel}, q_\parallel, T_\parallel, u_\parallel$. 
Considering the gyrotropic limit, the closure for $\widetilde{r}_{\perp\perp}$ defined as $r_{\perp\perp}=2p_\perp^2/\rho+\widetilde{r}_{\perp\perp}$ is simply $\widetilde{r}_{\perp\perp}=0$.
The $\widetilde{r}_{\parallel\perp}$ is defined as $r_{\parallel\perp}=p_\parallel p_\perp/\rho+\widetilde{r}_{\parallel\perp}$, and introducing for brevity
$\mathcal{T}_\perp \equiv \frac{T_\perp^{(1)}}{T_\perp^{(0)}} + \Big(\frac{T_\perp^{(0)}}{T_\parallel^{(0)}}-1\Big)\frac{B_z}{B_0}$,
there are 2 static closures, for
$R_{1}(\zeta)$:  $q_\perp^{(1)} =-\frac{p_\perp^{(0)}}{\sqrt{\pi}}  v_{\textrm{th}\parallel} \mathcal{H}\mathcal{T}_\perp$, and 
for $R_{2,0}(\zeta)$: $\widetilde{r}_{\parallel\perp}^{(1)} = -\frac{\sqrt{\pi}}{2} v_{\textrm{th}\parallel} \mathcal{H} q_\perp^{(1)}$,
which up to replacing $B_z$ with $|\boldsymbol{B}|$ (that comes here from a complete linearization),
are equivalent to the closures of \cite{Snyder1997}. 
There are also 6 dynamic closures, some of them related.   
With 3-pole approximants, a closure can be constructed for
$R_{3,1}(\zeta)$:
$\big[\frac{d}{dt}- \frac{\sqrt{\pi}}{4-\pi} v_{\textrm{th}\parallel} \pr_z \mathcal{H} \big] \widetilde{r}_{\parallel\perp}^{(1)}
   = -v_{\textrm{th}\parallel}^2 \frac{\pi}{2(4-\pi)} \pr_z q_\perp^{(1)}$,
and for $R_{3,2}(\zeta)$:
$\big[\frac{d}{dt}- \frac{2}{\sqrt{\pi}}v_{\textrm{th}\parallel} \pr_z \mathcal{H}\big] \widetilde{r}_{\parallel\perp}^{(1)} = -v_{\textrm{th}\parallel}^2 \pr_z q_\perp^{(1)}$,
that in the vanishing Larmor radius limit are equivalent to closures of \cite{PassotSulem2007}. Here we report on a new closure that is constructed with $R_{3,0}(\zeta)$:
\begin{eqnarray}
&& \Big[\frac{d}{dt}-\frac{(3\pi-8)}{2\sqrt{\pi}(\pi-3)} v_{\textrm{th}\parallel} \pr_z \mathcal{H} \Big] \widetilde{r}_{\parallel\perp}^{(1)} = 
  -v_{\textrm{th}\parallel}^2 \frac{4-\pi}{2(\pi-3)} \pr_z q_\perp^{(1)} \nn\\
&& \quad - p_{\perp}^{(0)}v_{\textrm{th}\parallel}^3
  \frac{(16-5\pi)}{4\sqrt{\pi}(\pi-3)} \pr_z \mathcal{H}\mathcal{T}_\perp,
\end{eqnarray}
that has a higher p.s.p., $o(\zeta^3)$. No closures with 4-pole (or higher) approximants are possible for $\widetilde{r}_{\parallel\perp}$.  

Returning to a 1D geometry and considering closures at higher-order moments $X_n=m\int (v-u)^n f dv$,  
the closure for $X_5$ with the highest p.s.p., $o(\zeta^5)$, is constructed with $R_{6,4}(\zeta)$, and  
reads
\begin{eqnarray}
&&  \Big[ \frac{d}{dt}  - \frac{3(180\pi^2-1197\pi+1984)\sqrt{\pi}}{(801\pi^2-5124\pi+8192)} v_{\textrm{th}} \pr_z\mathcal{H} \Big] X_5^{(1)} \nn\\
&&  =  - v_\textrm{th}^2 \frac{3(675\pi^2-4728\pi+8192)}{2(801\pi^2-5124\pi+8192)} \pr_z \widetilde{r}^{(1)} \nn\\
&&  +  v_\textrm{th}^3 \frac{3(285\pi-896)\sqrt{\pi}}{2(801\pi^2-5124\pi+8192)} \pr_z\mathcal{H} q^{(1)} \nn\\
&&  - v_\textrm{th}^4 n_0  \frac{3(945\pi^2-8184\pi+16384)}{4(801\pi^2-5124\pi+8192)} \pr_z T^{(1)} \nn\\
&&  +  v_\textrm{th}^3 n_0 T_0 \frac{9(450\pi^2-2799\pi+4352)\sqrt{\pi}}{(801\pi^2-5124\pi+8192)}  \pr_z\mathcal{H} u^{(1)}. \label{eq:R64_closure}
\end{eqnarray}
The closure is plotted in Figure 1 as the orange dotted line. 
  Going higher in the fluid hierarchy, and decomposing $X_6=15p^3/\rho^2+\widetilde{X}_6$, the closure with the highest p.s.p., $o(\zeta^6)$,
is obtained with $R_{7,5}(\zeta)$, being
\begin{eqnarray}
&&  \Big[ \frac{d}{dt}  + \alpha_{x_6} v_{\textrm{th}}\pr_z\mathcal{H} \Big] \widetilde{X}_6^{(1)} 
  =  + \alpha_{x_5} v_{\textrm{th}}^2  \pr_z X_5^{(1)} \nn\\
&& \quad + \alpha_{r}  v_{\textrm{th}}^3 \pr_z\mathcal{H} \widetilde{r}^{(1)} 
  + \alpha_q v_{\textrm{th}}^4 \pr_z  q^{(1)}  \nn \\
&& \quad + \alpha_T v_{\textrm{th}}^5 n_0 \pr_z\mathcal{H} T^{(1)} 
   + \alpha_u  v_{\textrm{th}}^4 n_0 T_0 \pr_z u^{(1)}, 
\end{eqnarray}
with coefficients
\begin{eqnarray}
  \alpha_{x_6} &=& 18(1545\pi^2-9743\pi+15360)\sqrt{\pi}/D;\nn\\
  \alpha_{x_5} &=& 3(52425\pi^2-331584\pi+524288)/(2D);\nn\\
  \alpha_{r}  &=&  3(7875\pi^2-50490\pi+80896)\sqrt{\pi}/D;\nn\\
  \alpha_{q}  &=& 3(162000\pi^3-1758825\pi^2+6263040\pi \nn\\
               && \quad  -7340032)/(4D);\nn\\
  \alpha_{T} &=&  - 27(15825\pi^2-99260\pi+155648)\sqrt{\pi}/(2D);\nn\\
  \alpha_{u} &=& 3(189000\pi^3-1612215\pi^2+4534656\pi \nn\\
               && \quad -4194304)/(2D);\nn\\
          D &=& (10800\pi^3-120915\pi^2+440160\pi-524288).
\end{eqnarray}
The closure is plotted in Figure 1 as the red line.

The remarkable result that the reliable closures reproduce the exact kinetic dispersion relation (1) once $R(\zeta)$ is replaced by  $R_{n,n'}(\zeta)$ leads us
to conjecture that there exist reliable fluid closures that can be constructed for even higher moments, i.e. satisfying (1), once $R(\zeta)$ is replaced
by the $R_{n,n'}(\zeta)$ approximant. Furthermore, for a given n-th order fluid moment, the reliable closure with the highest power series precision is the
dynamic closure constructed with $R_{n+1,n-1}(\zeta)$. 
Indeed, for higher order fluid moments one should be able to construct closures with higher order $R_{n+1,n-1}(\zeta)$ approximants
that will converge to $R(\zeta)$ with increasing precision. Thus, one can reproduce linear Landau damping in the fluid framework to any desired precision,
which establishes the convergence of fluid and collisionless kinetic descriptions.

The convergence was shown here in 1D geometry for the example of a long-wavelength low-frequency ion-acoustic mode.
  Nevertheless, the 1D closures have general validity, i.e. from the largest astrophysical scales to the Debye length, and are of course
  valid also for the Langmuir mode.
However, there are limitations in modeling the Langmuir mode, since for $\kpar\lambda_D <0.2$, Landau damping disappears very quickly, and some
closures show a small positive growth rate instead.  

The next logical step would be to establish an analytic convergence of fluid and kinetic descriptions in a 3D geometry
in the gyrotropic limit. However, in 3D, for a given n-th order tensor $\boldsymbol{X}_n$, the number of its gyrotropic moments is equal to $1+\textrm{int}[n/2]$
and increases with $n$. Therefore, it might be more difficult to show the convergence in 3D, although the convergence should exist.

Concerning direct applicability of the derived closures, numerical simulations of turbulence show a peculiar behavior, in that at sub-proton scales,
the parallel velocity spectrum is always much steeper in kinetic simulations than Landau fluid simulations (e.g. Fig. 7 of \citep{PerronePassot2018}).
The $r_{\parallel\parallel}$ closure of \cite{HammettPerkins1990}, does not include coupling with the parallel velocity component, whereas
our new closures do and could explain the discrepancy.

Finally, to emphasize the importance of the closures obtained, consider 1-fluid models in 1D geometry with $\kpar\lambda_D\ll 1$, closed by a simple
Maxwellian (non-Landau fluid) closures $X_n=0$, for $n$ odd, $n\ge 3$; and $X_n=(n-1)!!\frac{p^{n/2}}{\rho^{n/2-1}}$, for $n$ even, $n\ge 4$ (or
that the deviation $\widetilde{X}_n=0$ for $n$ even). 
It can be shown by induction that the dispersion relation reads
\begin{eqnarray}
&&  n=\textrm{odd:} \quad \zeta^{n-1} -\frac{n!!}{2^{(n-1)/2}}=0;\nn\\
&&  n=\textrm{even:}\quad \zeta^{n}-\frac{(n-1)!!}{2^{n/2}}\big( n\zeta^2-\frac{n}{2}+1 \big)=0. \label{eq:bum}
\end{eqnarray}  
For $n=3$ the solution is $\zeta=\pm \sqrt{3/2}$, and $n=4$ yields $\zeta=\pm \sqrt{3/2+\sqrt{3/2}}$, $\zeta=\pm \sqrt{3/2-\sqrt{3/2}}$.
However, $n=5$ yields $\zeta=\pm(\frac{15}{4})^{1/4}$; $\zeta=\pm i (\frac{15}{4})^{1/4}$, and    
$n=6$ yields $\zeta=\pm 0.58$; $\zeta=\pm 1.75$; $\zeta=\pm 1.87i$. In fact, for $n>4$, the solution of (\ref{eq:bum}) will always yield modes
that are unstable, and such fluid models can not be used for numerical simulations. The closure for $n=4$, $r=3p^2/\rho$, is sometimes called 
the ``normal'' closure \cite{ChustBelmont2006}. Here we conclude that the ``normal'' closure is actually the last non-Landau fluid closure, and that
beyond the 4th-order moment, Landau fluid closures are required.
\begin{acknowledgments}
  We acknowledge support of the NSF EPSCoR RII-Track-1 Cooperative Agreement OIA-1655280. ML thanks the Italian Space Agency for support under Grant 2015-039-R.O.
\end{acknowledgments}  
\hyphenation{Post-Script Sprin-ger}
%
\end{document}